%% file: paper.tex
\documentclass{spie}

\usepackage{microtype}
\usepackage[separate-uncertainty,binary-units]{siunitx}
\usepackage{graphicx}
\usepackage[hidelinks]{hyperref}
\usepackage{booktabs}
\usepackage{fullpage}
\usepackage{amssymb}

\usepackage{tikz} 
\usepackage{tikz-timing} 
\usetikzlibrary{matrix,chains,trees,shadows,fit}

\usepackage[hang,perpage]{footmisc}

\makeatletter
\newcommand\authorcount{19}
\renewcommand\AB@authnote[1]{\ifnum\value{authors}<\authorcount\relax,\fi\textsuperscript{\normalfont#1}}

\makeatother

\begin{document}

\title{Control and systems software for the\\Cosmology Large Angular Scale Surveyor (CLASS)}

\author[a]{\href{https://orcid.org/0000-0002-4436-4215}{Matthew~A.~Petroff}}
\author[a]{\href{https://orcid.org/0000-0002-8412-630X}{John~W.~Appel}}
\author[a]{\href{https://orcid.org/0000-0001-8839-7206}{Charles~L.~Bennett}}
\author[a]{Michael~K.~Brewer}
\author[a]{Manwei~Chan}
\author[b]{\href{https://orcid.org/0000-0003-0016-0533}{David~T.~Chuss}}
\author[a]{Joseph~Cleary}
\author[a]{\href{https://orcid.org/0000-0002-0552-3754}{Jullianna~Denes~Couto}}
\author[c,a]{\href{https://orcid.org/0000-0002-1708-5464}{Sumit~Dahal}}
\author[a]{\href{https://orcid.org/0000-0001-6976-180X}{Joseph~R.~Eimer}}
\author[c,a]{\href{https://orcid.org/0000-0002-4782-3851}{Thomas~Essinger-Hileman}}
\author[d]{\href{https://orcid.org/0000-0002-2061-0063}{Pedro~Flux\'a~Rojas}}
\author[e,a]{\href{https://orcid.org/0000-0003-1248-9563}{Kathleen~Harrington}}
\author[f,a]{\href{https://orcid.org/0000-0001-7466-0317}{Jeffrey~Iuliano}}
\author[a]{\href{https://orcid.org/0000-0003-4496-6520}{Tobias~A.~Marriage}}
\author[a,c]{Nathan~J. Miller}
\author[a]{\href{https://orcid.org/0000-0003-3487-2811}{Deniz~Augusto~Nunes~Valle}}
\author[g,a]{\href{https://orcid.org/0000-0002-5437-6121}{Duncan~J.~Watts}}
\author[h,a]{\href{https://orcid.org/0000-0001-5112-2567}{Zhilei~Xu}}
\affil[a]{Department of Physics \& Astronomy, Johns Hopkins University, Baltimore, MD 21218, USA}
\affil[b]{Department of Physics, Villanova University, Villanova, PA 19085, USA}
\affil[c]{NASA Goddard Space Flight Center, Greenbelt, MD 20771, USA}
\affil[d]{Facultad de F\'isica, Pontificia Universidad Cat\'olica de Chile, 7820436 Macul, Santiago, Chile}
\affil[e]{Department of Astronomy \& Astrophysics, University of Chicago, Chicago, IL 60637, USA}
\affil[f]{Department of Physics \& Astronomy, University of Pennsylvania, Philadelphia, PA 19104, USA}
\affil[g]{Institute of Theoretical Astrophysics, University of Oslo, 0315 Oslo, Norway}
\affil[h]{MIT Kavli Institute, Massachusetts Institute of Technology, Cambridge, MA 02139, USA}

\authorinfo{Further author information: (Send correspondence to M.~A. Petroff)\\E-mail: petroff@jhu.edu}

\maketitle

\begin{abstract}
The Cosmology Large Angular Scale Surveyor (CLASS) is an array of polarization-sensitive millimeter wave telescopes that observes $\sim{}70\%$ of the sky at frequency bands centered near \SI{40}{\giga\hertz}, \SI{90}{\giga\hertz}, \SI{150}{\giga\hertz}, and \SI{220}{\giga\hertz} from the Atacama desert of northern Chile. Here, we describe the architecture of the software used to control the telescopes, acquire data from the various instruments, schedule observations, monitor the status of the instruments and observations, create archival data packages, and transfer data packages to North America for analysis. The computer and network architecture of the CLASS observing site is also briefly discussed. This software and architecture has been in use since 2016, operating the telescopes day and night throughout the year, and has proven successful in fulfilling its design goals.
\end{abstract}


\section{Introduction}

Polarization anisotropy in the cosmic microwave background (CMB) provide a wealth of information about the early Universe. The Cosmology Large Angular Scale Surveyor (CLASS) is an array of polarization-sensitive millimeter wave telescopes that observes $\sim{}70\%$ of the sky at frequency bands centered near \SI{40}{\giga\hertz}, \SI{90}{\giga\hertz}, \SI{150}{\giga\hertz}, and \SI{220}{\giga\hertz} from the Atacama desert of northern Chile.\cite{Essinger-Hileman2014, Harrington2016} The primary science goals of CLASS are two-fold---to search for polarization B-modes\cite{Kamionkowski1997} as strong evidence of cosmological inflation and to make a measurement of the optical depth due to reionization, via polarization E-modes, with a sensitivity that is near the cosmic variance limit.\cite{Watts2015,Watts2018} The inflation paradigm postulates that the Universe underwent a period of exponential expansion in its first moments. Such inflationary expansion would produce tensor perturbations in the form of gravitational waves, which would leave an imprint on the CMB.\cite{Guth1981} The photons of the CMB were emitted during the period of decoupling that followed the epoch of recombination. Photons streaming through space collided with these free electrons, producing linear polarization via Thomson scattering. In the presence of the tensor perturbations induced by gravitational waves, this scattering produces B-mode polarization in addition to E-mode polarization, which is additionally produced by scattering in the presence of the dominant scalar perturbations. Thus, B-mode polarization in the CMB is evidence for gravitational wave-induced quadrupolar fluctuations during recombination, which is evidence for inflation. After a period known as the Dark Ages, the first stars formed, which reionized the neutral hydrogen in the Universe. This resulted in more Thomson scattering, which, in the presence of primordial gravitational waves, produces a second B-mode feature in the CMB, as well as additional scalar-sourced E-mode features.\cite{Hu1997} This polarization signal is at large angular scales, which CLASS is uniquely sensitive to among ground-based experiments.

CLASS, which has been successfully observing since 2016, seeks to accomplish its science goals by mapping the polarization of the CMB at large angular scales ($2 \lesssim \ell \lesssim 200$). Fast front-end polarization modulation, in the form of a variable-delay polarization modulator (VPM) as its first optical element,\cite{Chuss2012,Harrington2018} gives CLASS the stability necessary to measure the largest angular scales on the sky while also allowing most forms of instrument polarization to be rejected. Furthermore, the sky is observed at multiple frequencies to enable Galactic foreground signals, those from synchrotron and thermal dust in particular, to be disentangled from the primordial CMB signal.

In order to perform these measurements, software and computing and networking equipment are necessary to perform observations, collect time-ordered data, and package and transfer data for further analysis. To this end, this paper describes the architecture of the software used to control the telescopes, acquire data from the various instruments, schedule observations, monitor the status of the instruments and observations, create archival data packages, and transfer data packages to North America for analysis. The telescopes operate at a high-altitude site in northern Chile located on Cerro Toco and are connected to the internet via a facility in the nearby town of San Pedro de Atacama. Data are then transferred to a data server located on the Johns Hopkins University campus in Baltimore, Maryland for further analysis. The remainder of this paper is structured as follows. Section~\ref{sec:architecture} gives an overview of the network and software architecture, Section~\ref{sec:hardware} describes how this software is interfaced to some key subsystems, Section~\ref{sec:pipeline} details the data packaging and transfer pipeline, and Section~\ref{sec:web} outlines the web interface used to monitor status and schedule observations. Finally, we present some lessons learned in Section~\ref{sec:lessons} and conclude in Section~\ref{sec:conclusions}.

\section{Network and software architecture}
\label{sec:architecture}

With multiple receivers and numerous housekeeping systems, more than a dozen
computers are used to operate CLASS. Networking and a cohesive
software control suite are therefore required to coordinate the operation of these
systems such that science data can be collected, transferred off-site, and later
analyzed.

\subsection{Network layout}

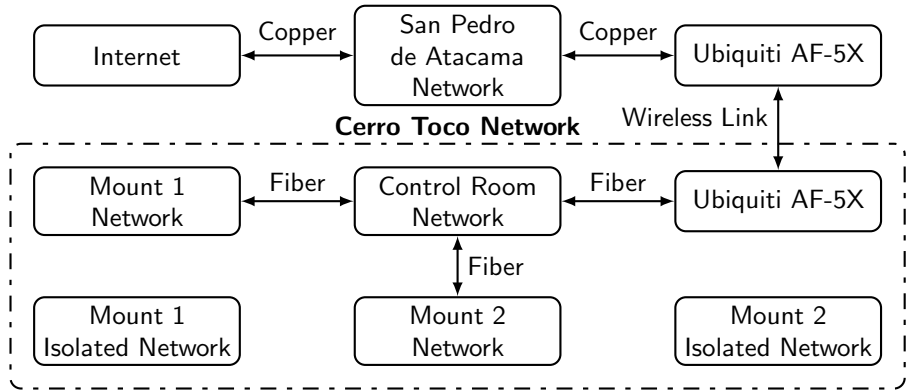
\begin{figure}
\centering
\input{network-diagram.tex}
\caption{Network architecture overview. Network segments in San Pedro de Atacama and on Cerro Toco are connected via a wireless link; other segments are connected via either copper or fiber optic Ethernet cables.}
\label{fig:network}
\end{figure}

The CLASS network is divided into two main components, the site network
on Cerro Toco and the supporting systems down the mountain in San Pedro de
Atacama. These components are connected using a wireless
network link operating at \SI{5}{\giga\hertz}. This approximately
\SI{43}{\km} line-of-sight radio link is provided by two Ubiquiti AF-5X
transceivers\footnote{Ubiquiti Inc.; \url{https://www.ui.com/}}
and is able to sustain data rates in excess of \SI{100}{Mbps}.
The site end of the wireless link is located on a tower \SI{\sim 170}{\meter} from the telescopes, at the edge of a cliff and with a clear line-of-sight to the San Pedro de Atacama end of the link. The site end of the link is powered via a solar panel and batteries and is connected to the rest of the site network via a fiber optic cable, providing complete electrical isolation, to protect against lightning damage.
All computers are
powered via uninterruptible power supplies (UPSs) so that they remain online during the brief power
losses sustained while switching between the site's two generators. Similarly, the equipment in San Pedro de Atacama is powered with a UPS; this UPS is capable of providing several hours of backup power, which is necessary due to the frequency and duration of power outages at this location. A overview of the CLASS network is given in Figure~\ref{fig:network}.

The San Pedro de Atacama portion of the network consists of a
router, an analysis computer, and a remote access computer.
The \textsc{analysis machine} is
used for doing preliminary analysis of data collected by the telescope.
Additionally, it is used to store telescope data once it is transferred
down the mountain, until the data are copied to North America.
The Cerro Toco portion of the network consists of three major segments: the
control room, the first mount, and the second mount. The \textsc{control room}
and \textsc{mounts} all use managed Ethernet switches, which provide gigabit connectivity over both copper and fiber optic cables, and are interconnected via fiber
optic cables. Each \textsc{mount} has an
additional switch for communication between the mount computers and the mount servo motors, which are
isolated from the primary network. The radio link to San Pedro de
Atacama is connected to the \textsc{control room} switch.
The \textsc{mounts} each contain two computers for controlling the mount; two
housekeeping computers, one for each cryostat; a star camera; 
four servo motors, which are on an isolated network; and two \textsc{VPM controllers}, one for each telescope.
The \textsc{control room} contains a server rack with a
\textsc{central server}, a display and status computer, a command terminal, a
star camera host computer, and detector readout computers, one for each
detector readout unit.\footnote{The detector readout units are on the mount and interface with the detectors; they are connected to the readout computers via fiber optic cables.} Additionally, there are laptop
computers that can be moved around as needed, to fulfill miscellaneous needs. A Wi-Fi access point is also installed, although this is turned off during observations.

\subsection{Software structure}

CLASS's software infrastructure is currently built around the Ubuntu~16.04
Linux distribution\footnote{Canonical Ltd.; \url{https://www.ubuntu.com/}} and
the Python programming language,\footnote{Python Software Foundation;
\url{https://www.python.org/}} specifically version~3.5. These releases are
both supported through at least 2021.\footnote{An upgrade to a newer version of Ubuntu is planned.} Software is developed using the
Git\footnote{\url{https://git-scm.com/}} distributed version control system.
All machines run the Ubuntu~16.04 operating system, except for the cryostat virtual machines
that run Windows~7 for compatibility with vendor-provided proprietary software and the mount
computers that run the VxWorks\footnote{Wind River Systems Inc.; \url{https://www.windriver.com/}}
real-time operating system (RTOS). With the exception of some C code required to interface with some
hardware and a few shell scripts, all control and data acquisition software running on the Linux servers is
written in Python. The control scripts are run as \texttt{systemd}\footnote{\url{https://www.freedesktop.org/wiki/Software/systemd/}} services so that they start on boot.

The basic software structure consists of Python scripts running each of the
telescopes' subsystems, with a central \textsc{command script} to coordinate
operations. Since control and data acquisition software is running on multiple computers
on the network simultaneously, a control and communications system for these
distributed systems is required. The \textsc{Pyro}\footnote{\url{https://pyro4.readthedocs.io/}} package, version~4, was chosen for this
purpose. Each subsystem is operated by a separate Python script, which includes a \textsc{Pyro} interface. These  \textsc{Pyro} interfaces expose the scripts' control function to the network. The \textsc{command script} is
then able to call these exposed functions as if they were local functions, with
\textsc{Pyro} seamlessly taking care of all of the network communications. Thus, subsystems are controlled by calling the \textsc{command script}, which in turn uses \textsc{Pyro} to call a function on the corresponding subsystem.

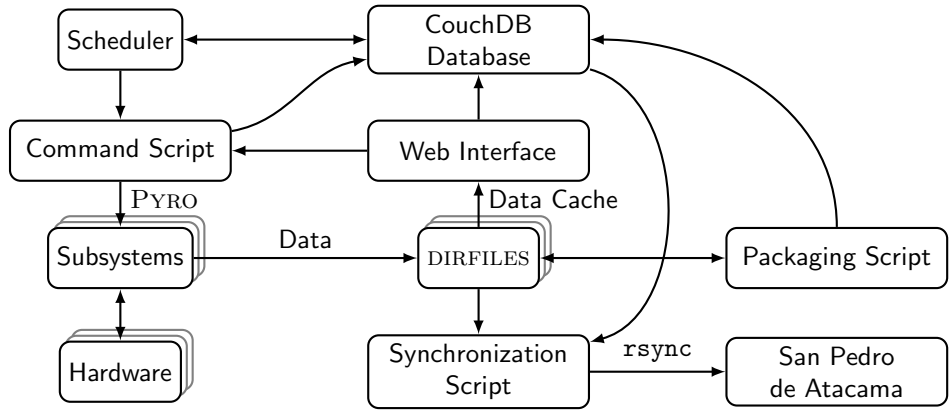
\begin{figure}
\centering
\input{software-structure.tex}
\caption{Software structure overview. Boxes represent various components or locations, and arrows represent data or command flow. The software is structured around Python scripts, with a web interface used for user interaction and \textsc{dirfiles} used for data recording.}
\label{fig:software-struct}
\end{figure}

Figure~\ref{fig:software-struct} contains an overview of the software
structure. All subsystems write data to the \textsc{central server} over the
network, and a centralized database is used to store metadata. A \textsc{scheduler} is used to execute preplanned observing
routines, and a web interface is used to monitor system statuses and control the
\textsc{scheduler}. The standard data format is the \textsc{dirfile} standard for
time-ordered data.\footnote{\url{http://getdata.sourceforge.net/dirfile.html}} This
filesystem-based, column-oriented format has been used for previous cosmology
experiments, including the Atacama Cosmology Telescope\cite{Switzer2008} and the EBEX\cite{Milligan2010} balloon-borne experiment. It is supported by its reference
\textsc{GetData} implementation\cite{GetData}, which includes the \texttt{pygetdata} Python bindings,
and can be easily visualized using the Kst plotting
software.\footnote{KDE e.V.; \url{https://kst-plot.kde.org/}}

\subsection{Scheduling}
\label{sec:scheduling}

While a central \textsc{command script} is sufficient, and preferable, for
testing, its use for regular observations is highly inefficient and error-prone.
Instead, a scheduling system is needed so an observation can be planned
in advance. For a survey experiment like CLASS, once a reasonable
observation strategy is found and tested, it will be used again and again.
Therefore, a scheduling system that facilitates running such preplanned
observations needs to be devised.
One way to implement such a system would be to replace the \textsc{command
script} with a scheduler, calling the same \textsc{Pyro} functions as the command
script. Although this method has the appeal of being simple and
straightforward, it has the disadvantage of possibly producing different
behavior than the \textsc{command script}, which would lead to much confusion
once a schedule is created using commands that were first tested by hand using
the \textsc{command script}. Therefore, the CLASS \textsc{scheduler}
calls the \textsc{command script}, to ensure the exact same
behavior.\footnote{More specifically, it calls a symbolic link to the
\textsc{command script}. The \textsc{command script} uses this difference in
the name of the script called to log the fact that the call was made by the
\textsc{scheduler}. There is also an additional mode that evaluates
the command arguments but does not execute anything, which allows command syntax
to be verified.} The results of the commands are logged to a database. A
schedule entry is created using the arguments to the \textsc{command script},
exactly what would be typed into the command terminal.
These entries are combined with lines containing a simple time delta syntax,
relative to either the schedule's start time or the last time directive, as
well as optional comment lines. An example of the \textsc{scheduler} syntax can
be seen in Figure~\ref{fig:syntax}.

\begin{figure}
\centering
\setlength{\fboxsep}{12pt}
\fbox{\begin{minipage}{0.8\textwidth}
\emph{\# This is a comment}\newline
\texttt{MCEQ -autobias}\newline

\emph{\# The following sets the commands after it to be run 2 minutes}\newline
\emph{\# after the last time directive}\newline
\texttt{/+2m}\newline
\texttt{MCEQ -acqdata 50000000}\newline

\emph{\# The following sets the commands after it to be run}\newline
\emph{\# 2 days, 20 hours, 5 minutes, 30 seconds after the start time of the schedule}\newline
\texttt{/2d20h5m30s}\newline
\texttt{MCEQ -stop}
\end{minipage}}
\caption{Example of \textsc{Scheduler} syntax. Examples are shown of individual commands, comments, and relative and absolute time deltas.}
\label{fig:syntax}
\end{figure}

\subsection{Databases}

While the science data are written to disk, it is helpful to keep various
sorts of metadata in a database for easy access and manipulation. For
CLASS, metadata about chunks of science data are stored in a database as
well as a log of all commands executed and \textsc{scheduler} data.
Metadata for the science data collected are entered into a database to
facilitate both data transfer and analysis. Therefore, a distributed database
system is needed such that it can be accessed locally at the telescope site, in
San Pedro de Atacama, and in North America. Three desirable
properties for distributed computer systems, in this case a distributed
database, are consistency, availability, and partition tolerance. However,
Brewer's theorem states that only two of the three can be achieved by any one
system \cite{cap}. Consistency requires all copies of the database to return
identical results when queried, availability requires the database to always be
accessible, and partition tolerance requires the system to still function even
if the network link between the different copies is severed. For example, if a
database system enforces both partition tolerance and consistency, the database
will be read-only, and thus only partially accessible, if the network link goes
down, since the link is needed to keep the different copies consistent. For
CLASS, availability and partition tolerance are the most important
properties, since both the radio link to the telescope site and the internet
connection to North America might be unreliable, but this should not stop data
collection, which also involves adding metadata to the database. Although this
could in general lead to merging problems when the database system tries to restore
consistency once a severed network link is reestablished, the software running at CLASS's three
locations modify different database fields, so no conflicts will arise. Most
data are entered at the telescope site, with the software at the two auxiliary sites only
modifying fields pertaining to their local data storage locations.

To this end, the CouchDB database\footnote{Apache Software Foundation;
\url{https://couchdb.apache.org/}} was chosen for handling metadata, with a server at each
location running a database server instance that hosts a full copy of the
databases to allow for fast, low-latency database queries. CouchDB is
a NoSQL database that stores data as JSON documents and
is highly available and partition tolerant, with eventual consistency. NoSQL
databases differ from traditional SQL databases by not requiring a fixed,
predefined table structure; instead, each document can have any data keys it
needs, functioning more like a dictionary than a spreadsheet. Instead of using
structured queries to retrieve data, mapping and reduction functions are
written and used; in the CouchDB parlance, these are known as ``views'' and are
written in JavaScript.

The CouchDB instance hosts databases to store commands executed, schedules,
the results of commands executed for schedules, and
metadata about the science data products. The \textsc{commands database} records
commands executed by the \textsc{command script} along with a timestamp and Git
revision hashes for both the \textsc{command script} and the target script;
this logging happens regardless of how the \textsc{command script} was
called---directly or by the \textsc{scheduler}. Each commit in a Git repository
has a unique 160-bit SHA-1 hash \cite{sha1} associated with it; since all of
the site software is contained in a Git repository, recording these revision
hashes uniquely identifies the exact version of the scripts that were executed,
for future reference. The \textsc{schedules database} contains schedules for
use by the \textsc{scheduler}. Each entry contains the actual schedule entered,
a parsed copy of the schedule for use by the \textsc{scheduler}, an ID number,
descriptive tags, the schedule's author, a start time, and a status field. Once
the schedule finishes running, or fails, an end time is added. Failed schedules
also record the line that failed. These entries are created by the
\textsc{scheduler}'s web interface and are modified by it and the actual
\textsc{scheduler}. The \textsc{scheduler} also records the results of commands
executed in a \textsc{jobs database}; each entry contains the command executed,
a timestamp, the command's output, and if it failed. The final
database contains metadata for each packaged chunk of CLASS science
data.
In addition to this standard metadata, which are also kept with the data packages,
the database stores the location of the data packages at each site: Cerro Toco, San
Pedro de Atacama, and North America. Each site only modifies the location field
for itself, so synchronization issues after a network outage are avoided. These
location fields are used to facilitate data transfer and analysis.

\section{Hardware interfaces}
\label{sec:hardware}

To capture data, a physical instrument is required. To record these data and to
control these instruments and other hardware, computer interfaces to the
hardware are required, as is software to use these hardware interfaces.
Although the hardware of different CLASS subsystems differ considerably,
their software interfaces were designed to be similar, using a Python script that interfaces either directly with the hardware or through vendor-provided software to expose a \textsc{Pyro} control interface. Where custom electronics are used, such as for cryogenic diode-based thermometry, warm thermometry in the telescopes' optics cages, and for control of the telescopes' VPMs, these are designed around an Ethernet interface. While TCP is used for sending commands, UDP is used to stream data in a manner that is robust to connectivity interruptions and equipment restarts. For equipment that uses an RS-232 serial interface, such as AC resistance bridges (used for reading out ruthenium oxide cryogenic thermometers), magnetometers, the diesel tank level sensor, and the site weather station, Ethernet-to-serial converters are used to allow access over the site network; this simplifies operations, since the control scripts for these equipment can be run from the \textsc{central server}.\footnote{A USB-to-serial interface is used instead for the weather station, since it is mounted on the outside of the control room, a short distance from the \textsc{central server}.} Along with scientific instruments, other site infrastructure, such as the generators, can also be controlled remotely.

The only exceptions to this general control architecture are the cryostats. They are controlled using a proprietary GUI software provided by their manufacturer, Bluefors,\footnote{Bluefors Oy; \url{https://bluefors.com/}} which only runs under the Windows operating system and does not provide much in the way of a scriptable interface. Therefore, the cryostats are operated manually, with the \textsc{Remote Desktop Protocol} used to access the GUI. However, as the cryostats only need to be controlled when they are being cooled down or warmed up for maintenance, not during normal operation, this lack of integration with the rest of the site operations software does not impede observations.

\subsection{Detectors}

The telescopes' transition edge sensor detectors\cite{Denis2009, Dahal2020} are read out via time-division multiplexing using Multi-Channel Electronics (MCE) units\cite{Battistelli2008}
developed at the University of British Columbia.\footnote{\url{https://e-mode.phas.ubc.ca/mcewiki/index.php/Main_Page}} There is one MCE per receiver and currently one or two receivers per telescope mount, although the two mounts will eventually have two receivers each. The detectors are
connected to the MCE using a set of SQUID multiplexers at
the focal plane and a SQUID amplifier series array at the \SI{4}{\kelvin} stage of the cryostat.
The MCE is triggered using a \textsc{frame pulse}, delivered with a clock
signal over fiber optics from a \textsc{sync box}, one per telescope mount, which is a device used to synchronize data collection
by multiple MCEs and the telescope mount. The \textsc{frame pulse}
assigns a 32-bit ID number to the readout frame, which consists of a full focal
plane readout, and delivers pulses to trigger reading
out each detector. The \textsc{frame pulse} is delivered at
slightly over \SI{200}{\hertz}.\footnote{Running
slightly fast is preferable to running slightly slow, since it ensures the
recorded data chunks are less than ten minutes in length. This simplifies the
data packaging process, since it guarantees that a data chunk only needs to be
split once.} The MCE, mounted on the cryostat, communicates with a
\textsc{host computer} in the \textsc{control room} via fiber optics and a PCI
readout card. This \textsc{host computer} uses
\textsc{MCE acquisition software (MAS) scripts}, provided by the University of British Columbia, to
control the MCE and record data. A \textsc{Pyro} interface allows
use of these scripts over the network and also provides a routine for measuring $I$--$V$ response and applying appropriate detector biasing. Data are written to an NFS mount on the \textsc{central server}.

\subsection{Mount}

Each telescope mount is controlled by an industrial \textsc{mount computer}
running the VxWorks RTOS, which loads its software over the network at boot time. The \textsc{mount computer}, also known as the antenna control unit (ACU),
controls the mount's four servo motors, two for azimuth and one
each for elevation and boresight angle, over an isolated Ethernet network and also reads in various encoders and tiltmeters. The \textsc{mount computer} also receives \textsc{frame pulse}
information from the \textsc{sync box} over RS-485 (instead of the fiber optics used for the MCEs) for synchronization with the
MCE units and has a GPS receiver to obtain precise timing
information; as the MCEs only record \textsc{frame pulse} information and not time, the mount's timing information is also used to assign times to detector data when the data are packaged. The mount's position is determined using encoders that feed a pointing model,
and these data, as well as timing and status information, are written
to an NFS mount on the \textsc{central server}. The pointing model is derived from Moon and planet observations \cite{Xu2020}, although the initial pointing model was constructed using star camera observations processed using the Astrometry.net blind astrometric solver \cite{Lang2010}.\footnote{The star camera consists of a networked machine vision camera in a waterproof enclosure that is rigidly attached to the telescope mount.} The \textsc{mount computer} exposes a Telnet
interface and a TCP interface to the site network. The Telnet interface
allows for interactive control via a terminal and status monitoring, while the TCP interface is used for scripted control of the
mount. A \textsc{Pyro} interface running on the \textsc{central server} translates
this TCP interface into a standard \textsc{Pyro} interface for the
\textsc{command script}.

The mount monitor and control system, which is written in C++, implements the programs used for scanning in azimuth, with sun avoidance
to preclude the boresight from coming within \SI{20}{\deg} of the Sun; sky dips, for atmospheric
analysis; and drift scans of planets and the Moon, which are used for pointing and beam analysis.
The JPL DE430 ephemerides\cite{DE430} are used by the monitor and control system for tracking the Moon, Sun, and
planets. Parameters used by these programs are passed to the monitor and control system via the \textsc{Pyro} interface.

\subsection{Variable-delay polarization modulator}

The VPM consists of a fixed wire array and a movable mirror, to modulate
the incoming polarization signal. The mirror is driven by a closed-loop motion control
system. This system consists of voice coils to actuate the mirror and encoders
to measure its position. The encoder signals are duplicated, with one copy going to
the \textsc{VPM controller} and another copy going to the ACU. The \textsc{VPM controller}, built around an industrial applications processor and running a network-booted RTOS, uses this information to maintain proper
mirror position, while the ACU logs the encoder
data. This duplication is necessary because the \textsc{VPM controller} is not connected to the \textsc{sync box}, and the VPM encoder positions need to be recorded synchronously with the detector data. A control script running on the \textsc{central server} communicates
with and configures this \textsc{controller} via TCP over an Ethernet
connection. Finally, a \textsc{Pyro} interface exposes a standard network
interface for VPM control.

\section{Data pipeline}
\label{sec:pipeline}

In order to measure the polarization of the CMB, data need to be
acquired from the telescopes' detectors and various housekeeping systems,
combined, packaged, and transferred to North America for analysis. This
packaging needs to happen in real time, to prevent a backlog and maximize
observing time; furthermore, the process should be lossless and verifiable to
prevent data loss and identify potential data corruption.

\subsection{Acquisition}

Data acquisition starts with scheduling an observation and acquiring data from hardware instruments. As data are acquired, they are written as \textsc{dirfiles}
on the \textsc{central server}, either directly or using NFS mounts. These are structured with a directory for each computer acquiring data, with
subdirectories for each acquisition system. Individual \textsc{dirfiles} are named using
a timestamp formatted as \texttt{\%Y-\%m-\%d-\%H-\%M-\%S}, e.g.,
\texttt{2020-03-22-20-00-00}. \textsc{Dirfiles} consist of a directory that contains a plain-text
\textsc{format file} and separate binary files for each stream of time-ordered data.
Additional derived fields can be defined as linear combinations of existing
fields, e.g., to scale raw data to SI units; aliases, e.g., to label the
location of thermometers without having to edit the acquisition code; and
extracted bit-fields, e.g., to combine multiple status flags in one raw data
field.

Data are divided into ten-minute chunks as they are acquired. For asynchronous
housekeeping data, which are not acquired using the \textsc{sync box}, a new
chunk is started at clock-aligned intervals, e.g., at 00:00, 00:10, 00:20, etc., so the first data chunk in the series will be less than ten
minutes in length, e.g., six minutes if data collection started at twenty-four
minutes past the hour. These divisions are done when the time in seconds since
the Unix Epoch\footnote{\texttt{1970-01-01T00:00:00Z}} modulo 600 seconds rolls
over. Synchronous data are acquired in arbitrarily-aligned chunks of
approximately ten minutes in length.\footnote{The chunks are \num{120000} frame
pulses long, or just under ten minutes, ``redefining'' the second as 200
\textsc{frame pulses}.} All timestamps are in Universal Time, with asynchronous
data using the computer's NTP-synchronized system clock in UTC and
the mount using UT1, derived from GPS time; the detector data are not
acquired with timestamps, but they are synchronized with the mount data using the
\textsc{sync box}. The time is stored in the \textsc{dirfiles} as seconds since the
Epoch.

\subsection{Packaging}

Once data are acquired, they must be packaged into a standard data product. The
standard CLASS data product is a clock-aligned ten-minute \textsc{data
chunk} consisting of time-ordered data, with one \textsc{dirfile} per mount for
synchronous data and individual \textsc{dirfiles} for each asynchronous data acquisition
system. Synchronous data are combined in this data product, since this can be
done without altering any data or losing any information. Since the \textsc{dirfile}
format keeps each data field in a separate file and the \textsc{GetData} library does not
load the data until a specific field is read, this combining of data from
multiple receivers does not add any overhead when only one receiver's data
need to be read. Asynchronous data are kept separate
due to the losses inherent to resampling and interpolation. This is a trade-off
between making things simpler during analysis and preventing data loss, since
interpolation eventually needs to be done. A packaging script is run every ten minutes, at
00:02, 00:12, 00:22, etc. It processes data from two data periods
prior and before, e.g., the 00:32 run will only process data from the 00:10
period and before; this ensures that the non-ten-minute-boundary-aligned
synchronous data collection for the data period has finished. For example,
synchronous data collection might start at 00:19, which would include data in
the 00:10 period; this collection would not end until 00:29, unlike the
asynchronous data collection, which would start a new data chunk at 00:20.
Therefore, the maximum delay, plus a few minutes of buffer, is used.

\subsubsection{Synchronous}

The detector and ACU data sampling on a given mount is synchronized, but the two mounts are not synchronized with each other. An overview of the synchronous data packaging process can be seen in
Figure~\ref{fig:synchronous}.
The first step in packaging synchronous data is to divide them into CLASS's
standard ten-minute data intervals. This process starts with dividing each data
chunk collected by the ACU into two parts, \texttt{A} and \texttt{B}, where
seconds since the Epoch modulo 600 seconds rolls over; the ACU data must be the first processed, since they are
the only synchronous data that contain timestamps, instead of just \textsc{frame pulse} numbers. These divided chunks are
then combined to form ten-minute chunks aligned with the Epoch, with zero-padding added before, after, and between chunks when needed
to form a full ten-minute chunk. The field containing the \textsc{frame pulse}
number is filled in with the appropriate values instead of zero-padding like
the rest of the fields; this allows for easier synchronization with the
detector data. A status field is also added to the resulting
\textsc{dirfile} that shows where the ACU data are valid, as opposed to where they are
zero-padded, stored in an 8-bit integer field.

Next, the detector data from each MCE unit on the mount are divided using
the \textsc{frame pulse} numbers recorded in both the MCE data and the now aligned ACU data. As
with the ACU data, these chunks are then combined to form aligned ten-minute
chunks, again with zero-padding added before, after, and between chunks as
needed; again, a status field is added, but different bits in the 8-bit integer
field are used. Care must be taken, since the \textsc{frame pulse} counter
occasionally rolls over.\footnote{As the \textsc{frame pulse} is a \SI{32}{\bit}
unsigned integer, this happens roughly every eight months. This counter also resets any time the \textsc{sync box} is power-cycled.} Finally, the raw
data files in each of the MCE \textsc{dirfiles} are moved
into subdirectories in the ACU \textsc{dirfile}. The ACU \textsc{dirfile} \textsc{format file} is altered to include the MCE \textsc{dirfile} \textsc{format files} as child fragments, which prefix the MCE \textsc{dirfile} field names with the appropriate MCE name, and the data
valid status fields are combined using a bitwise \textsc{or}. After this
process is complete, the resulting \textsc{dirfile} can be treated as an asynchronous
data chunk in the next step of the packaging process.

The merged data are then compressed, with \texttt{gzip} compression used for the mostly floating-point ACU data and FLAC compression\footnote{Xiph.Org Foundation; \url{https://xiph.org/flac/}} used for the fixed-precision MCE data. As FLAC was designed as an audio compression codec, it does not support the 32-bit data recorded by the MCE, but \textsc{GetData} avoids this limitation by splitting each 32-bit integer into two 16-bit integers and compressing the data as two separate channels. However, this is still not optimal, since FLAC will treat these integers as smaller than 16~bits if the most significant bits are never used. CLASS operates the MCEs in \textsc{data mode~10}, which uses the lower 25~bits of the 32-bit integer to store low-pass-filtered detector data as a signed integer and the upper 7~bits to store a flux jump counter as another signed integer. Thus, when the flux jump counter is zero or positive, the upper half of the 32-bit integer is treated as having fewer than 16~bits, often as having as few as 9--10~bits, but is treated as having the full 16~bits when the flux jump counter is negative. Therefore, the data packaging script removes the 7-bit flux jump counter from the combined data stream and converts it to a separate 8-bit signed integer field, which is FLAC-compressed. The filtered detector data that remain in the original 32-bit field is then treated as a 25-bit integer when FLAC compression is applied. This scheme performs well on the CLASS data, reducing disk space usage while still providing rapid decompression and data access.

\begin{figure}
\centering
\input{synchronous-packaging.tex}
\caption{Overview of synchronous data packaging process. The procedure by which ACU (mount) and MCE (detector) data are aligned and joined is shown.}
\label{fig:synchronous}
\end{figure}

\subsubsection{Asynchronous}

\begin{figure}
\centering
\input{data-product}
\caption{Overview of the CLASS data product structure. Solid boxes represent
directories, dashed boxes represent \textsc{dirfiles}, and dotted boxes represent
other files. Only a subset of the \textsc{dirfiles} and images are shown. The subdirectories in the \texttt{async} directory contain asynchronous data associated with a specific receiver, while the \texttt{sync.zip} \textsc{dirfile} combines ACU data for a given mount with detector data from both of its receivers.}
\label{fig:data-structure}
\end{figure}
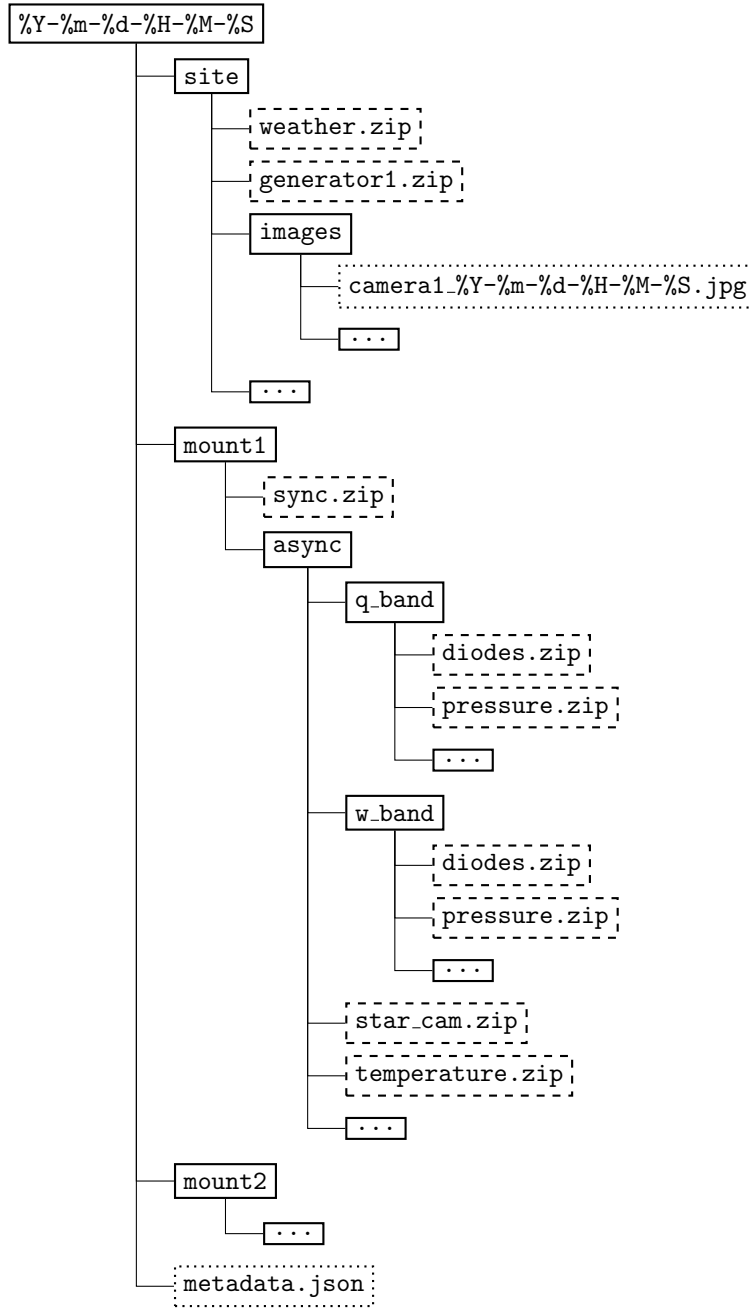

Asynchronous data are recorded to timestamped \textsc{dirfiles} in a directory structure
based on the computer recording the data. However, the final data product
structure starts with a timestamped directory, with subdirectories for
different types of data, a hierarchy divorced from the computer systems used to
collect the data. Figure~\ref{fig:data-structure} contains an overview of this
structure. Therefore, the first step of the asynchronous data packaging
process, which includes the merged synchronous data, rearranges the
asynchronous \textsc{dirfiles} from their original structure to the final data product
structure, in a new location. As the vendor-provided software for operating the cryostats only records CSV files,
these data are converted to \textsc{dirfiles} at this step. While the \textsc{dirfiles} are originally recorded uncompressed, at this stage they are compressed using the \texttt{gzip} compression format to save space.\footnote{FLAC compression is not used for the asynchronous data, since many of the data fields involve floating point numbers, which FLAC compression does not support.} Images from each of the telescope site's monitoring cameras are also packaged, since this provides a method of checking for bad weather or unusual site activity should artifacts be found in the telescope data during analysis.

Next, the \textsc{dirfiles} are converted to uncompressed \textsc{zip} file\footnote{ISO/IEC 21320-1:2015} archives. An uncompressed \textsc{zip} file concatenates all of the files it contains together and then includes a file offset table that allows for its contents to be located and read in a manner that allows for random reads. Encapsulating the \textsc{dirfile} into a \textsc{zip} archive reduces the number of files in the data package by more than an order of magnitude, which provides significant speed improvements for data transfer and backup operations, due to the small size of many of the files contained in the unencapsulated \textsc{dirfiles}.\footnote{Unfortunately, this extended \textsc{dirfile} functionality is only available as a patch, since the \textsc{GetData} library seems to be abandoned. The author of this manuscript (M.A.P.) has not received a response from the \textsc{GetData} maintainer (D.V.W.) with either postings to the \texttt{getdata-devel@lists.sourceforge.net} mailing list or via private correspondence in more than two years, nor has there been any public activity on the project in that time frame.} Finally, a
JSON metadata file is created that contains basic information about the
\textsc{data chunk}, including the files included, their sizes, and SHA-1
checksums, for later verification; this file is saved in the root of the
\textsc{data chunk}. The location of the \textsc{data chunk} on disk is then
appended to the metadata information, and this information is saved to the
database.

\subsection{Transfer}

Once the data are packaged, they need to be transferred off the mountain to San
Pedro de Atacama and then to North America. Data integrity needs to be
verified after each transfer, and old data eventually need to be deleted from the \textsc{central server} to
free up disk space. Data are transferred from the telescope site's \textsc{central server} to the
\textsc{analysis machine} in San Pedro de Atacama via the CLASS network's
radio link. A  synchronization script runs on the site's \textsc{central
server} every ten minutes, which transfers the data files using
\texttt{rsync}.\footnote{\url{https://rsync.samba.org/}}
This script uses a lock file to ensure only one copy of itself is running at a
time, and a certificate is used to authenticate with the \textsc{analysis
machine}. A verification script is also run every ten minutes on the
\textsc{analysis machine} in San Pedro de Atacama. This script checks each unverified
\textsc{data chunk} on the machine against the local copy of the database and
sees if the \textsc{data chunk} has a local path assigned to it. If it does not
have a local path, the \textsc{data chunk} is verified against the database,
and the \textsc{chunk}'s local path is added to the database. If the
verification fails, an error is thrown, and the local path is not added to the
database, so it is treated as if the data were never transferred. The corrupted
data might be fixed by \texttt{rsync} with the next synchronization run---else,
manual intervention is required to delete the corrupted data. If the
\textsc{data chunk} already has a local path, the data have already been
verified, so the \textsc{chunk} is ignored. Since the synchronization script
uses \texttt{rsync}'s delayed updates feature, files do not appear until the
transfer is complete, so issues with incomplete transfers are mitigated.
This process is repeated to transfer data from the \textsc{analysis machine} in San Pedro de Atacama
to a data server located in Baltimore.

\subsubsection{Archiving of data and deletion of old data}

While a permanent copy
of the data will be kept on the file server in Baltimore, the data will not be permanently kept on disk
on-site or in San Pedro de Atacama, as disk space will need to be freed for new
observations. On both the \textsc{central server} on Cerro Toco and the
\textsc{analysis machine} in San Pedro de Atacama, a script is run on a
daily basis to free up space by removing old \textsc{data chunks} that have
already been transferred and verified, to ensure a minimum amount of free disk
space; as long as enough space is free, data will not be deleted. When disk
space needs to be freed, the database will be checked for \textsc{data chunks}
that have already been transferred to San Pedro de Atacama in the case of the
telescope site's \textsc{central server} and transferred to North America in the case of the
\textsc{analysis machine}. Of these \textsc{data chunks}, data will be deleted
in chronological order, with the oldest data first, until enough free disk
space is available; currently, this threshold is set at 75\% of the disk capacity. If this process is not able to free enough space, an error will be
thrown, and manual intervention will be required. The only existing copy of a
\textsc{data chunk} will never be automatically deleted.

\subsubsection{Ensuring data integrity}

A number of steps are taken throughout the data packaging and transfer process
to ensure data integrity. At each step in the process, data are not deleted
until after the next step in the process has finished. Updating the database
with a \textsc{data chunk}'s location at a new location is always the last step
in the process, to avoid the possibility of the database showing that the data
are there when they are not actually there; if something goes wrong in the transfer
process, some or all of the data might be there, but the database will be
conservative and list the data as not being present.

As the last step in the packaging procedure, SHA-1 checksums are created for
each file and stored in the database. These checksums are unique, but
repeatable, 160-bit cryptographic hashes based on the data; if even one bit in
the source file changes, a completely different hash will be generated, so
transfer errors of either the data or the checksums are easily detected. After
each transfer step, new checksums are generated for each file, and these
checksums are compared to the copies stored in the database. If the checksums
match, the transfer was successful. If they don't match, the transfer failed,
and data corruption occurred; in this case, the data are treated as if they were
not transferred at all, and an error is thrown. Along with being replicated
across multiple locations, regular offline backups are made of the CLASS
CouchDB database that stores the science data product metadata.

The primary CLASS data server located in Baltimore utilizes a ZFS file system.\footnote{\url{https://openzfs.org/}} ZFS provides strong data integrity guarantees, including against silent data corruption, by making extensive use of checksums and through regular integrity checking. A \texttt{RAIDZ2} configuration is used to protect against disk failures, and regular file system snapshots are taken to protect against inadvertent data deletion, caused either by human error or malware. Nightly backups are made to a replica data server, located in a different location, to provide additional protection. These backups rely on ZFS's ability to efficiently send and receive file system snapshots and are configured in such a way that even if one of the two servers is compromised by a malicious actor, it would be difficult to erase both copies of the data. Finally, the data packages are additionally backed up to a cloud storage provider. As of November 2020, CLASS has \SI{16}{\tera\byte} of compressed data packages recorded to disk.

\section{Web interface}
\label{sec:web}

A web interface is run on the CLASS site's
\textsc{central server} to provide status information, facilitate execution of commands and scheduling of
observations, and provide a wiki to store site operations information and
documentation. Except for the wiki, which uses the MediaWiki software package,\footnote{Wikimedia Foundation; \url{https://www.mediawiki.org/}} the interface runs using Python and the
Django web framework\footnote{Django Software Foundation;
\url{https://www.djangoproject.com/}} and uses the Bootstrap\footnote{\url{https://getbootstrap.com/}} front-end framework.

\begin{figure}
\centering
\frame{\includegraphics[width=0.95\textwidth]{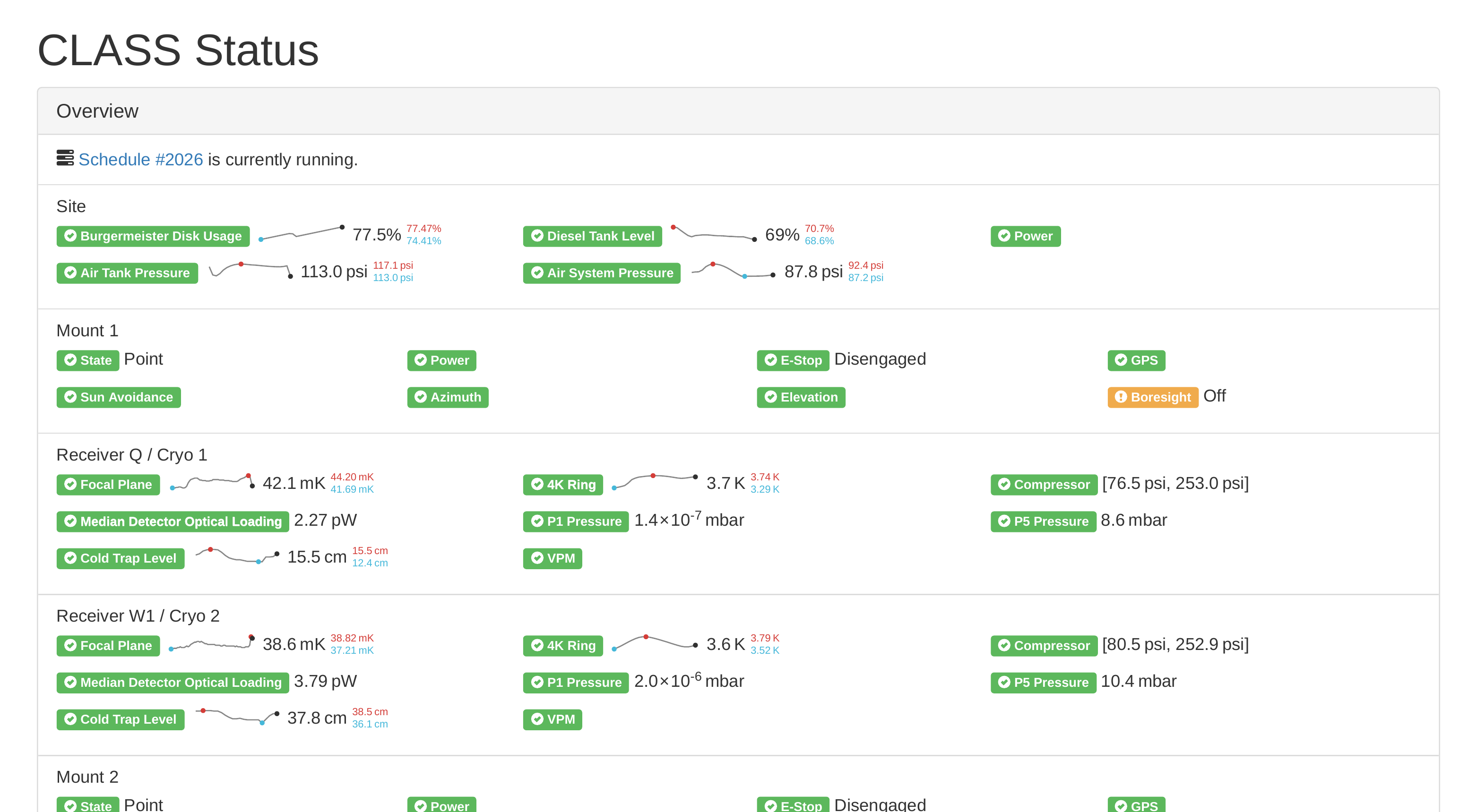}}
\caption{Web status interface. This screenshot shows an example view from the web status interface overview section, which allows one to quickly evaluate instrument health.}
\label{fig:status-display}
\end{figure}

The status display provides pertinent status information about the telescope
systems, including cryostat temperatures, pressures, and flows; available disk
space; and site environmental conditions.
A screenshot of the status display overview section can be seen in Figure~\ref{fig:status-display}, which provides a color-coded dashboard for quickly evaluating the current instrument health as well as sparklines\cite{Tufte2006} for evaluating how parameters have changed over the past day.
In addition to the overview, detailed pages are available for each cryostat and mount. Here, data are graphed client-side
using a JavaScript charting library; multiple time ranges can be selected, and real-time
updates are pushed using a WebSockets connection for the shortest time range. A
script continuously reads in housekeeping data from the currently recording
\textsc{dirfiles} and stores it in a day-long circular cache; this cache is flushed to
disk every ten minutes, so it will still contain data if the caching script
needs to be restarted. This caching script exposes a \textsc{Pyro} interface,
which the web interface back-end uses to access the data. The script
additionally triggers the pushing of the real-time updates over the WebSockets
connection. Due to the packaging process, only the currently recording \textsc{dirfile}
is accessible, necessitating the cache, which is also useful to improve performance. Although the web interface back-end
could directly cache the data, this is not done, since the web server runs
multiple, simultaneous copies of the back-end to maintain responsiveness; this
would cause duplicate disk access and, more importantly, cause duplicate
real-time updates to be pushed. In addition to providing information for the status display, the caching script also sends alerts\footnote{These are currently sent to a channel in the \emph{Slack} chat tool used by the CLASS collaboration, although the destination of the alerts can be easily changed.} for certain failures that require immediate attention, such as a cryostat warming up, although in some cases alerts are directly triggered by the corresponding hardware interface script instead. The status display also provides access to results from automated preliminary analysis scripts, which are run every morning on the \textsc{analysis machine} in San Pedro de Atacama, and to current views from the various site observation cameras, either as static images updated once per minute or as live video.

\begin{figure}
\centering
\frame{\includegraphics[width=0.95\textwidth]{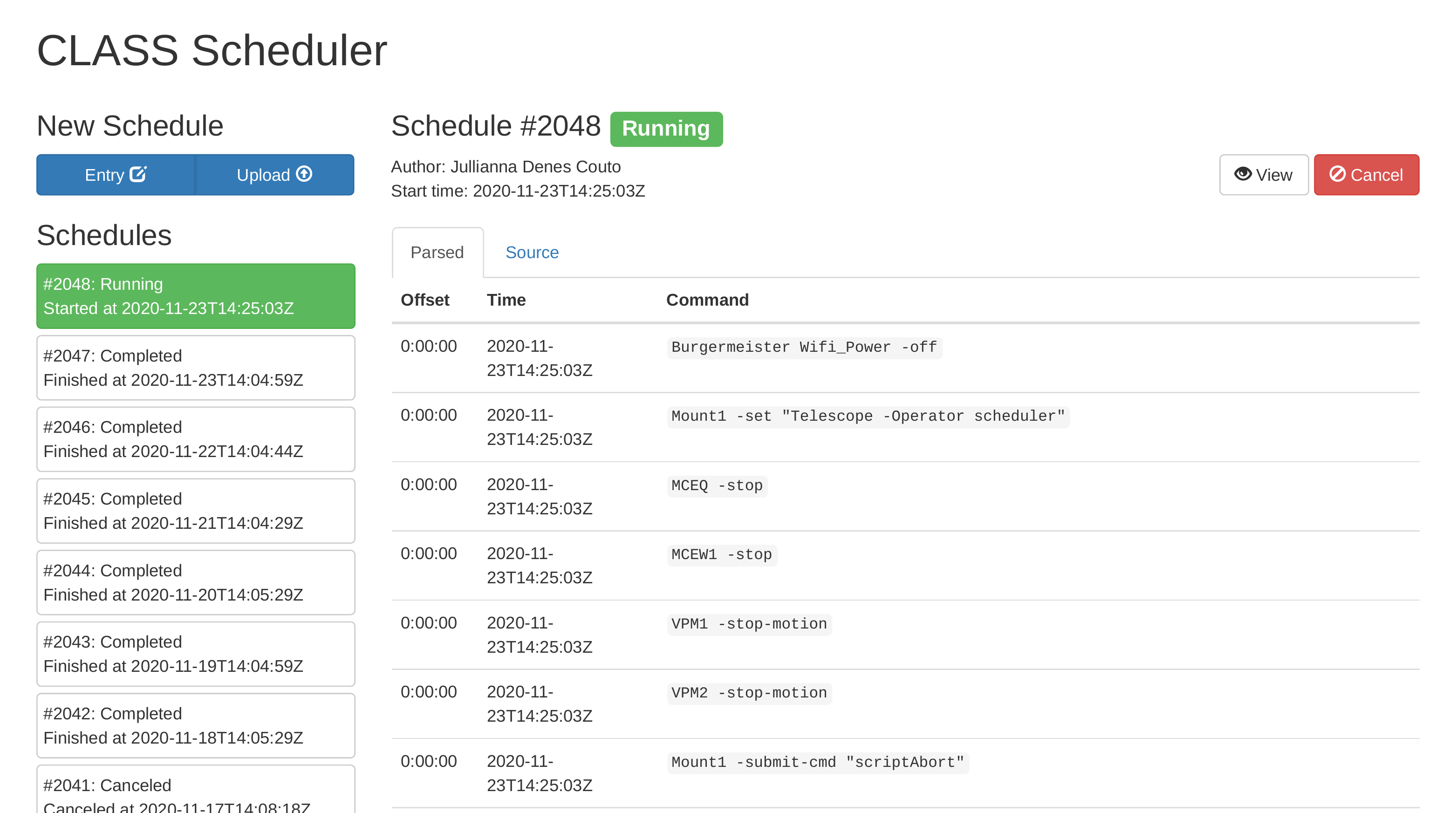}}
\caption{Web scheduler interface main screen. The main screen of the web scheduler interface is shown in this screenshot, displaying the currently running schedule.}
\label{fig:scheduler-main}
\end{figure}

\begin{figure}
\centering
\frame{\includegraphics[width=0.95\textwidth]{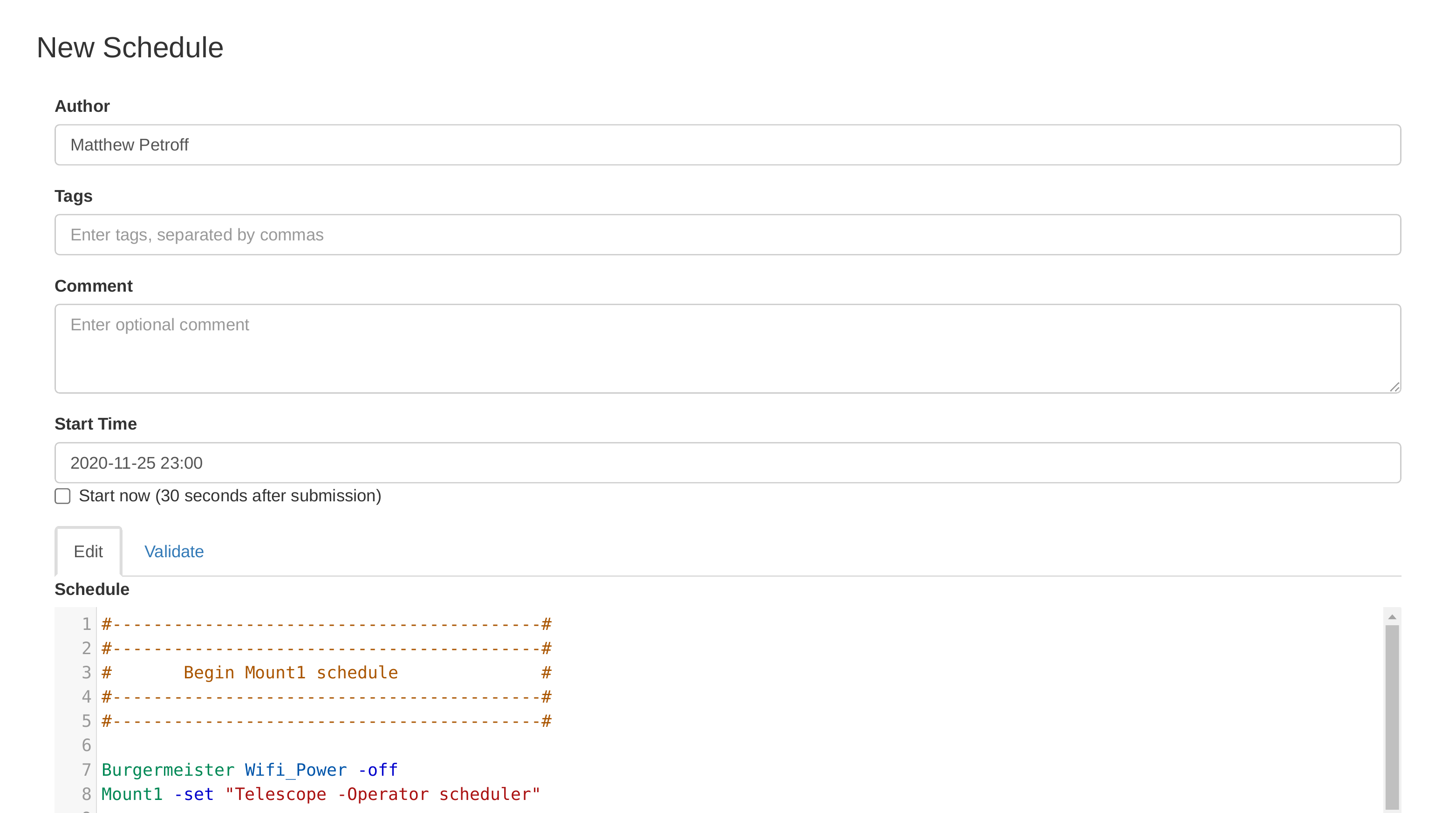}}
\caption{Web scheduler new schedule interface. This screenshot shows a new schedule being written; metadata can be added, the start time can be adjusted, and the schedule can be validated before it is submitted to be run.}
\label{fig:scheduler-add}
\end{figure}

The \textsc{scheduler} interface provides an easy-to-use, graphical method for
scheduling observations on the telescopes. Login credentials are required to
access and use the \textsc{scheduler} interface. The main screen shows the
currently running schedule if one is running, the next pending schedule if no schedule is
currently running but one is pending, or the last schedule run if none are currently running or pending. Figure~\ref{fig:scheduler-main} shows this interface. A sidebar
provides a list of recent schedules and their status information, with a link
to a complete history of schedules. When viewed, each schedule can be displayed
either in the raw form it was entered in, with basic syntax highlighting, or in
a parsed form. Status, start time, title, author, schedule number, and tagging
information are also displayed. If a schedule completed, the end time is
shown. If a schedule failed to complete, the failure time and failed command
are displayed. Pending schedules can be edited, and pending or running
schedules can be canceled.
An interface for adding new schedules is also accessible from the main screen.
This prompts a user to enter metadata including a title, start time, and,
optionally, tags; a date and time picker is provided to assist the user with
start time entry. The author field is automatically filled in with the user's
name, but can also be edited.\footnote{The username of the user that entered the schedule is also recorded to the database as a separate field but is not displayed.} Then comes a box for entering the schedule; basic
syntax highlighting is provided. Any time in the editing process, the user can
switch between the edit box and a verification tab, which verifies the syntax
of the entered schedule; lines containing incorrect syntax are highlighted.
Once finished, the schedule can be submitted and will be run by the
\textsc{scheduler}. Previously-run schedules also provide a button for opening the new schedule interface with its fields pre-filled with data from the previous schedule. This interface for adding schedules can be seen in
Figure~\ref{fig:scheduler-add}. A separate interface is also provided for executing, and displaying the results of, commands outside the context of a schedule. This interface is normally locked out while a schedule is running to prevent accidental execution of commands, but it can be unlocked if necessary. Additionally, it displays a list of recently run commands and shows in real-time which other users have the command interface open, to encourage communication.

\section{Lessons learned}
\label{sec:lessons}

Some parts of the architecture described above were different when the first CLASS telescope was deployed in 2016, and other parts should have been done differently in hindsight or would be done differently without certain hardware interface restrictions. Other changes have been incremental updates, such as transitioning from Python~2 to Python~3.\footnote{Although Python~3 was fairly well established when software development started in 2014, Python 2 was originally used, since, at the time, the \textsc{GetData} library did not yet support Python~3.}
One lesson that was learned is that interfacing with instruments over Ethernet with a control script running on the \textsc{central server} is easier and more reliable than running a control script on another computer and writing data over NFS, since the former is a ``pull'' configuration, while the latter is a ``push'' configuration; ``pull'' configurations are less likely to run into issues if network connectivity is temporarily lost or if the \textsc{central server} needs to be rebooted. With this insight, thermometry data recording was transitioned from using USB interfaces to using Ethernet interfaces, which improved reliability. Ideally, this transition would be extended to its logical conclusion, using a single server to operate the telescopes and all the instruments. Unfortunately, limitations with the MCE interfaces and cryostat interfaces prevent this. MCE interface limitations additionally hamper improvements to data acquisition synchronization. Without these limitations, the detector, pointing, and VPM encoder data acquisition hardware could be configured to receive time information over the network via the precision time protocol standard, which allows for sub-microsecond time distribution. Data frames could then be synchronized with the start of GPS time seconds, as long as the data acquisition rate is limited to an integer number of samples per second, allowing for the elimination of the dedicated synchronization hardware that is currently used. Another lesson learned is that a large number of small files makes data transfer and backup operations much slower than they would otherwise be. This led to a transition from standard \textsc{dirfiles} to \textsc{dirfiles} encapsulated in \textsc{zip} files, which reduced the number of files stored on disk by more than an order of magnitude and significant performance improvements for data transfer and backup operations. In hindsight, it would have also likely been better if the data packages were based on hour-long chunks instead of ten-minute chunks for the same reason, but the current data package chunking scheme is too ingrained to change now for what would be a more incremental improvement.

\section{Conclusions}
\label{sec:conclusions}

The current CLASS control and systems software architecture has been described. This architecture is based around Python scripts, with \textsc{dirfiles} used to record data and the \textsc{Pyro} library used for network-based control interfaces. Data are packaged into ten-minute long chunks, which are aligned with the start of the hour. Scheduling and status monitoring are performed using a web interface. Throughout, there is a focus on ensuring data integrity.
It is our hope that our documenting of the CLASS software architecture here can help inform software development for future experiments. The CLASS software architecture has been successful at fulfilling its intended purpose and has proven to be sufficiently reliable.

\acknowledgments

We acknowledge the National Science Foundation Division of Astronomical Sciences for their support of CLASS under Grant Numbers 0959349, 1429236, 1636634, 1654494, and 2034400. The CLASS project employs detector technology developed in collaboration between JHU and Goddard Space Flight Center under several previous and ongoing NASA grants. Detector development work at JHU was funded by NASA grant number NNX14AB76A. ZX is supported by the Gordon and Betty Moore Foundation. We acknowledge scientific and engineering contributions from Max Abitbol, Mario Aguilar, Fletcher Boone, David Carcamo, Francisco Espinoza, Saianeesh Haridas, Connor Henley, Yunyang Li, Lindsay Lowry, Isu Ravi, Gary Rhodes, Daniel Swartz, Bingie Wang, Qinan Wang, Tiffany Wei, and Ziang Yan. We acknowledge productive collaboration with Dean Carpenter and the JHU Physical Sciences Machine Shop team. We thank Mar\'ia Jos\'e Amaral, Chantal Boisvert, William Deysher, and Miguel Angel D\'iaz for logistical support. We further acknowledge the very generous support of Jim and Heather Murren (JHU A\&S '88), Matthew Polk (JHU A\&S Physics BS '71), David Nicholson, and Michael Bloomberg (JHU Engineering '64). CLASS is located in the Parque Astron\'omico Atacama in northern Chile under the auspices of the Agencia Nacional de Investigaci\'on y Desarrollo (ANID). The software described in this manuscript makes extensive use of NumPy.\cite{NumPy}

\bibliography{paper.bib}
\bibliographystyle{spiebib}

\end{document}

%% file: network-diagram.tex
\begin{tikzpicture}
  \sffamily
  \tikzset{>=latex}
  \matrix (m) [matrix of nodes, 
    column sep=15mm,
    row sep=8mm,
    nodes={draw,
      thick,
      anchor=center, 
      text centered,
      rounded corners,
      minimum width=1.5cm, minimum height=8mm
    },
    txt/.style={text width=2.5cm}
  ] { 
    |[txt]| {Internet} 
  & |[txt]| {San Pedro de Atacama Network} 
  & |[txt]| {Ubiquiti AF-5X} 
  \\
    |[txt]| {Mount 1 \\ Network} 
  & |[txt]| {Control Room Network}  
  & |[txt]| {Ubiquiti AF-5X}  
  \\
    |[txt]| {Mount 1 \\ Isolated Network} 
  & |[txt]| {Mount 2 \\ Network}  
  & |[txt]| {Mount 2 \\ Isolated Network}  
  \\
  };

  { [start chain,every on chain/.style={join}, every join/.style={thick, <->}]
    \path[style={draw, thick, <->}] (m-1-1) edge node [above] {Copper} (m-1-2);
    \path[style={draw, thick, <->}] (m-1-2) edge node [above] {Copper} (m-1-3);
    \path[style={draw, thick, <->}] (m-1-3) edge node [left, yshift=1.5mm] {Wireless Link} (m-2-3);
    \path[style={draw, thick, <->}] (m-2-3) edge node [above] {Fiber} (m-2-2);
    \path[style={draw, thick, <->}] (m-2-2) edge node [above] {Fiber} (m-2-1);
    \path[style={draw, thick, <->}] (m-2-2) edge node [right] {Fiber} (m-3-2);
  };

  \tikzset{dotted/.style={draw=black, thick,
                          dash pattern=on 2pt off 4pt on 6pt off 4pt,
                          inner sep=3mm, rectangle, rounded corners}};

  \node (dotted box) [dotted, fit = (m-2-1) (m-3-3)] {\hspace*{\fill}~};

  \node at (dotted box.north) [above, inner sep=1mm] {\textbf{Cerro Toco Network}};
\end{tikzpicture}

%% file: software-structure.tex
\begin{tikzpicture}
  \sffamily
  \tikzset{>=latex}
  \matrix (m) [matrix of nodes, 
    column sep=18mm,
    row sep=6mm,
    nodes={draw,
      thick,
      anchor=center, 
      text centered,
      rounded corners,
      minimum width=1.5cm, minimum height=8mm
    },
    txt/.style={text width=2.7cm},
    multi/.style={fill=white,double copy shadow={draw=gray}}
  ] { 
    Scheduler 
  & |[txt]| {CouchDB Database} 
  & 
  \\
    |[txt]| {Command Script} 
  & |[txt]| {Web Interface} 
  & 
  \\
    |[multi]| Subsystems 
  & |[multi]| \textsc{dirfiles} 
  & |[txt]| {Packaging Script} 
  \\
    |[multi]| Hardware 
  & |[txt]| {Synchronization Script}
  & |[txt]| {San Pedro de Atacama} 
  \\
  };

  { [start chain,every on chain/.style={join}, every join/.style={thick, <->}]
    \path[style={draw, thick, ->}] (m-2-1) edge[out=10,in=190] (m-1-2);
    \path[style={draw, thick, ->}] (m-2-2) edge (m-1-2);
    \path[style={draw, thick, ->}] (m-2-2) edge (m-2-1);
    \path[style={draw, thick, ->}] (m-1-1) edge (m-2-1);
    \path[style={draw, thick, <->}] (m-1-1) edge (m-1-2);
    \path[style={draw, thick, ->}] (m-3-2) edge node [right,yshift=2pt] {Data Cache} (m-2-2);
    \path[style={draw, thick, ->}] (m-2-1) edge node [right,yshift=2pt] {\textsc{Pyro}} (m-3-1);
    \path[style={draw, thick, ->}] (m-3-1) edge node [above] {Data} (m-3-2);
    \path[style={draw, thick, <->}] (m-3-3) edge (m-3-2);
    \path[style={draw, thick, ->}] (m-1-2) edge[out=-15,in=15] (m-4-2);
    \path[style={draw, thick, ->}] (m-3-3) edge[out=90,in=0] (m-1-2);
    \path[style={draw, thick, <->}] (m-3-1) edge (m-4-1);
    \path[style={draw, thick, ->}] (m-3-2) edge (m-4-2);
    \path[style={draw, thick, ->}] (m-4-2) edge node [above] {\texttt{rsync}} (m-4-3);
  };
\end{tikzpicture}

%% file: synchronous-packaging.tex
\begin{minipage}{5in}
\noindent The process starts with misaligned synchronous data.

\begin{center}
\begin{tikztimingtable}[%
    timing/dslope=0.1,
    timing/.style={x=2.5ex,y=2ex},
    x=2.5ex,
    timing/rowdist=3ex,
    timing/name/.style={font=\rmfamily\scriptsize}
]
Aligned Period & 10D{00:00} 10D{00:10} \\
ACU Data & 9D{A} 10D{B} 1D{C} \\
MCE Data & 3D{a} 10D{b} 7D{c}  \\
\extracode
\begin{pgfonlayer}{background}
\begin{scope}[semithick]
\vertlines[black,dotted]{0,2,...,20}
\vertlines[black]{0,10,20}
\end{scope}
\end{pgfonlayer}
\end{tikztimingtable} \\
\end{center}

\noindent ACU data are then split at the proper alignment based on timestamps...

\begin{center}
\begin{tikztimingtable}[%
    timing/dslope=0.1,
    timing/.style={x=2.5ex,y=2ex},
    x=2.5ex,
    timing/rowdist=3ex,
    timing/name/.style={font=\rmfamily\scriptsize}
]
Aligned Period & 10D{00:00} 10D{00:10} \\
ACU Data & 9D{A2} 1D{B1} 9D{B2} 1D{C1} \\
MCE Data & 3D{a} 10D{b} 7D{c}  \\
\extracode
\begin{pgfonlayer}{background}
\begin{scope}[semithick]
\vertlines[black,dotted]{0,2,...,20}
\vertlines[black]{0,10,20}
\end{scope}
\end{pgfonlayer}
\end{tikztimingtable} \\
\end{center}

\noindent ...and then joined to form aligned chunks.

\begin{center}
\begin{tikztimingtable}[%
    timing/dslope=0.1,
    timing/.style={x=2.5ex,y=2ex},
    x=2.5ex,
    timing/rowdist=3ex,
    timing/name/.style={font=\rmfamily\scriptsize}
]
Aligned Period & 10D{00:00} 10D{00:10} \\
ACU Data & 10D{A2 + B1} 10D{B2 + C1} \\
MCE Data & 3D{a} 10D{b} 7D{c}  \\
\extracode
\begin{pgfonlayer}{background}
\begin{scope}[semithick]
\vertlines[black,dotted]{0,2,...,20}
\vertlines[black]{0,10,20}
\end{scope}
\end{pgfonlayer}
\end{tikztimingtable} \\
\end{center}

\noindent MCE data are then split at the proper alignment based on \textsc{frame pulse} numbers in the ACU data...

\begin{center}
\begin{tikztimingtable}[%
    timing/dslope=0.1,
    timing/.style={x=2.5ex,y=2ex},
    x=2.5ex,
    timing/rowdist=3ex,
    timing/name/.style={font=\rmfamily\scriptsize}
]
Aligned Period & 10D{00:00} 10D{00:10} \\
ACU Data & 10D{A2 + B1} 10D{B2 + C1} \\
MCE Data & 3D{a2} 7D{b1} 3D{b2} 7D{c1}  \\
\extracode
\begin{pgfonlayer}{background}
\begin{scope}[semithick]
\vertlines[black,dotted]{0,2,...,20}
\vertlines[black]{0,10,20}
\end{scope}
\end{pgfonlayer}
\end{tikztimingtable} \\
\end{center}

\noindent ...and then joined to form aligned chunks.

\begin{center}
\begin{tikztimingtable}[%
    timing/dslope=0.1,
    timing/.style={x=2.5ex,y=2ex},
    x=2.5ex,
    timing/rowdist=3ex,
    timing/name/.style={font=\rmfamily\scriptsize}
]
Aligned Period & 10D{00:00} 10D{00:10} \\
ACU Data & 10D{A2 + B1} 10D{B2 + C1} \\
MCE Data & 10D{a2 + b1} 10D{b2 + c1}  \\
\extracode
\begin{pgfonlayer}{background}
\begin{scope}[semithick]
\vertlines[black,dotted]{0,2,...,20}
\vertlines[black]{0,10,20}
\end{scope}
\end{pgfonlayer}
\end{tikztimingtable}
\end{center}
\end{minipage}

%% file: data-product.tex
\tikzstyle{every node}=[draw=black,thick,anchor=west]
\tikzstyle{df}=[dashed]
\begin{tikzpicture}[%
  grow via three points={one child at (0.5,-0.7) and
  two children at (0.5,-0.7) and (0.5,-1.4)},
  edge from parent path={(\tikzparentnode.south) |- (\tikzchildnode.west)}]
  \node {\texttt{\%Y-\%m-\%d-\%H-\%M-\%S}}
    child { node {\texttt{site}}
      child { node [df] {\texttt{weather.zip}}}
      child { node [df] {\texttt{generator1.zip}}}
      child { node {\texttt{images}}
        child { node [dotted] {\texttt{camera1\_\%Y-\%m-\%d-\%H-\%M-\%S.jpg}}}
        child { node {\texttt{...}}}
      }
      child [missing] {}
      child [missing] {}
      child { node {\texttt{...}}}
    }
    child [missing] {}
    child [missing] {}
    child [missing] {}
    child [missing] {}
    child [missing] {}
    child [missing] {}
    child { node {\texttt{mount1}}
      child { node [df] {\texttt{sync.zip}}}
      child { node {\texttt{async}}
        child { node {\texttt{q\_band}}
          child { node [df] {\texttt{diodes.zip}}}
          child { node [df] {\texttt{pressure.zip}}}
          child { node {\texttt{...}}}
        }
        child [missing] {}
        child [missing] {}
        child [missing] {}
        child { node {\texttt{w\_band}}
          child { node [df] {\texttt{diodes.zip}}}
          child { node [df] {\texttt{pressure.zip}}}
          child { node {\texttt{...}}}
        }
        child [missing] {}
        child [missing] {}
        child [missing] {}
        child { node [df] {\texttt{star\_cam.zip}}}
        child { node [df] {\texttt{temperature.zip}}}
        child { node {\texttt{...}}}
      }
    }
    child [missing] {}
    child [missing] {}
    child [missing] {}
    child [missing] {}
    child [missing] {}
    child [missing] {}
    child [missing] {}
    child [missing] {}
    child [missing] {}
    child [missing] {}
    child [missing] {}
    child [missing] {}
    child [missing] {}
    child { node {\texttt{mount2}}
      child { node {\texttt{...}}}
    }
    child [missing] {}
    child { node [dotted] {\texttt{metadata.json}}};
\end{tikzpicture}